# High-power single-cycle THz emission from large-area photoconductive emitters at 400 kHz

MOHSEN KHALILI,[1,*] YICHENG WANG[1], STEPHAN WINNERL,[2] AND CLARA J. SARACENO[1,3]

[1]*Photonics and Ultrafast Laser Science, Ruhr-University Bochum, 44801 Bochum, Germany*
[2]*Helmholtz-Zentrum Dresden-Rossendorf, Institute of Ion Beam Physics and Materials Research, 01328 Dresden, Germany*
[3]*Research Center Chemical Science and Sustainability, University Alliance Ruhr, 44801 Bochum, Germany*
**mohsen.khalilikelaki@ruhr-uni-bochum.de*

**We report high average power THz emission from a GaAs-based large-area photoconductive emitter, excited by a commercial Yb-laser amplifier without any compression schemes, doubled to the green. The LAE is pumped at 11.4 W of green average power (515 nm) and 310 fs pulse duration at 400-kHz repetition rate. We obtain a maximum THz power of 6.7 mW with a spectrum extended up to 3 THz. Using electro-optic sampling (EOS) for detection, we measure a high dynamic range of 107 dB in a measurement time of 70 s. To the best of our knowledge, this is the highest power THz source so far demonstrated by photoconductive emitters, based on a simple and robust, commercially available Yb-laser.**

The generation of broadband terahertz (THz) radiation has gained considerable interest for various scientific and technological applications, particularly due to the unique properties of the THz frequency range. THz waves have distinctive characteristics, including their ability to penetrate a wide range of non-conducting materials making them highly suitable for imaging and non-destructive testing applications [1,2]. Additionally, THz frequencies correspond to various vibrational, rotational, and low-energy electronic transitions in molecules, enabling spectroscopic analysis of complex microscopic systems [3,4]. However, many THz applications rely on achieving a high signal-to-noise ratio within short measurement times, which in the case of broadband, pulsed THz sources calls for highest possible average powers and repetition rate compatibility with fast acquisition methods [5].

Many efforts have been made in recent years to increase average power levels; most of which have been successful using optical rectification in nonlinear crystals. Current record values approach the watt-level average power, For instance, a tilted pulse front scheme in bulk lithium niobate and gas-plasma based THz source have achieved THz power above 640 mW [6,7], though they come with increased phase-matching complexity and their efficient operation still demands complex high energy laser systems. On the other hand, photoconductive (PC) emitters based on linear acceleration of photoinduced carriers in semiconductors using an external bias voltage [8,9], are well known to significantly decrease the demands for pulse energy. Instead, in PC emitters, the energy for carrier acceleration is supplied by the bias voltage, improving conversion efficiency without strongly depending on the optical pulse energy. State-of-the-art photoconductive antennas (PCAs) exhibit high THz conversion efficiency, achieving 3.4% in Rhodium-doped InGaAs emitters under an optical power of 28 mW on a 25 μm gap, generating 637 μW of THz power and reaching a dynamic range (DR) of 111 dB [10]. More recently, PCAs based on this material system have demonstrated THz powers of 970 μW with 55 mW of optical power on a 50 μm gap, achieving a DR of 101 dB [11]. Advances with iron-doped InGaAs materials have further pushed performance limits, achieving THz powers of 958 μW with 56.5 mW of pump power on a 50 μm gap and a record DR 137 dB over a 100-minute integration time, or 117 dB within 60 s [12].

A comparison of these results indicates that increasing the PC gap and optical power are promising but underexplored strategies for scaling THz output and dynamic range. While larger PC areas could theoretically enable greater power output in PCA structures, achieving this would require bias voltages in the kV range, presenting practical challenges [13,14]. This problem can be circumvented with microstructured inter-digitated large-area photoconductive THz emitters (LAE) [15–17], operating on the same principles as PCAs but their large active area in the range of 1-100 mm$^2$ allows for the use of higher average-power ultrafast driving lasers and moderate bias voltages of only a few volts [14,17]. Recently, we reported the first systematic investigation of LAE excitation at MHz repetition rates and high average power [18], which showed that thermal load from optical average power is the primary obstacle to THz power scaling at these very high repetition rates.

Another path which we explore here is the use of amplifier-based high-power sources operating at hundreds of kHz, providing higher single pulse energies while maintaining high average power and sufficient repetition rates for fast scanning and averaging. Conceptually, this scheme replaces averaging over many weak pulses by averaging over less, stronger individual THz pulses. If the conversion efficiency to the THz is significantly higher, this is advantageous over using much higher repetition rates, and fast averaging can still be achieved at hundreds of kHz using continuous scanning delays. So far, one early result explored this approach with a high-repetition rate Ti:Sapphire laser, with a maximum reported 1.5 mW of THz power at 250 kHz of repetition rate [15]. Further power scaling was here prevented by the limited average power of the Ti:Sapphire laser.

Here we report efficient, high-power THz generation from a 10×10 mm² LAE based on GaAs, excited by a frequency-doubled Yb-based commercial laser amplifier, which delivers up to 11.4 W average power at 515 nm. The LAE achieves a 6.7 mW THz output at 55 $V_{pp}$ bias voltage, yielding a conversion efficiency of $4.2 \times 10^{-3}$. A focused THz electric field of 31 kV/cm is obtained. Using EOS, a dynamic range of 107 dB is achieved within a 3 THz bandwidth. This simple, Yb-laser-driven source offers robust potential for THz imaging and spectroscopy applications.

The experimental setup for THz generation and detection is shown in Fig. 1a. A (Trumpf TruMicro 2000) at 1030 nm that delivers up to 17.5 W of average power at 400 kHz with 310 fs of pulse duration is employed as driving source. Notably, the amplifier is not optimized for achieving shortest pulse duration, as its spectral FWHM bandwidth is 8 nm, corresponding to a time-bandwidth product (TBP) of 0.70. To enable photon absorption in GaAs (bandgap energy 1.42 eV), we frequency-double the pump laser by a 300-μm BBO crystal, thus generating up to 11.4 W of green power at central wavelength 515 nm (photon energy 2.4 eV). The spectral bandwidth in the green is 2 nm, Although the autocorrelator device used in the setup cannot measure the green pulse duration directly, assuming the same TBP as the driving laser, the green pulse duration is estimated to be approximately 310 fs. This is a reasonable estimation given the small thickness of the BBO crystal and the rather long pulses used for doubling. The divergent green beam is enlarged by using lens $L_2$ before being focused by lens ($L_3$ = 50 mm) towards the LAE.

The LAE used for our experiment is based on semi-insulating-GaAs (SI-GaAs) emitter with an interdigitated electrode metal-semiconductor-metal (MSM) structure, which is more detailed in [14,19]. Fig. 1b shows the geometry of microstructure LAE. The MSM structure is masked by a second metallization layer isolated from the MSM electrodes. The isolation layer on MSM blocks the excitation laser beam in every second period of the MSM structure. In this way, charge carriers are excited and accelerated unidirectionally and radiated THz fields interferes constructively in the far field. The emitter is housed in a water-cooled copper mount (Fig. 1c) that extracts a small fraction of the accumulated heat from the emitter sides. The electrode structure, with a width and spacing of 10 μm, can produce a bias field on the order of several 10 kV/cm at a moderate bias voltage in the range of tens of volts. A waveform generator in combination with a voltage amplifier is used to provide a flexible voltage source with different waveshape and modulation frequency to use as bias voltage of the LAE. The LAE is placed at normal incidence after the focus of the lens and by moving the LAE in direction of the beam propagation can control the green spot size on it. Apart from small deviations due to diffraction that are relevant mainly for the longer wavelengths, the generated THz beam maintains the phase front of the incident pump beam [15]. The THz pulse is collimated and refocused for detection by using a pair of gold-coated OAP mirrors with focal lengths of 101.6 mm and 50.8 mm, respectively. The THz electric field is characterized using EOS [21] in combination with standard balanced detection scheme [22] and 3-mm gallium phosphide (GaP) as a detection crystal. A 1% output-coupler directs a small fraction of the laser power to the probe arm, where a fast-oscillating delay line (10 Hz, 15 ps range) periodically delays the probe beam relative to the THz beam. The THz power is measured by replacing the detection crystal with a pyroelectric power meter (Ophir, 3A-P-THz) and its performance was cross-checked to be identical with a calibrated THz power meter (SLT, calibrated by the German Metrology Institute PTB). The bias voltage signal synchronizes the lock-in amplifier for low-noise detection. To prevent pump leakage, filters (180 μm white paper and 0.5 mm high-density polyethylene) are used in the collimated THz arm and on the power meter. Filter transmission is measured and accounted for in power values. When the bias voltage or pump power is switching-off, THz power drops to the noise level, ensuring accurate THz power measurements without pump leakage or electrical heating from the LAE.

To find the optimal condition for the LAE excitation, we first investigate the dependence of the emitted THz power on optical power as a function of pump spot size on the LAE. A bipolar fully symmetric square shape voltage with 150 kHz of modulation frequency is used as bias voltage.

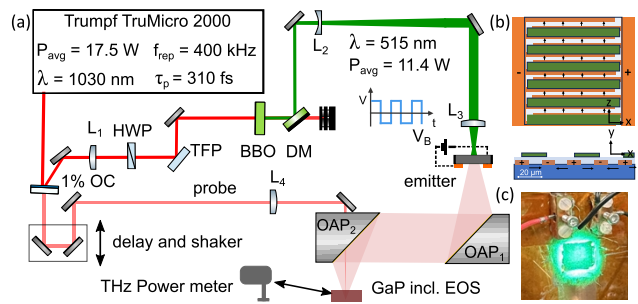

Fig. 1. (a) Experimental setup for THz generation and detection. HWP: half-wave plate, TFP: thin film polarizer. DM: dichroic mirror. (b) geometry of microstructure LAE (c) LAE in water-cooled copper mount.

In this measurement, the bias voltage is kept at 25 $V_{pp}$ to minimize the electrical thermal load [18]. Fig. 2a shows the saturation curves, while the excitation spot diameter ($1/e^2$) is swept from 1.9 mm to 9 mm. A maximum optical power of 11.4 W is safely applied for all the spot sizes without permanent damage on the LAE, although a clear power saturation observed for the smaller spot sizes. Increasing the spot size results in higher THz power, particularly at higher pump powers. In Fig. 2b we show THz conversion efficiency versus pump power. Increasing the spot size on the LAE results in higher THz power, particularly at higher pump powers. This improvement is attributed to 1) better heat extraction from the substrate, and 2) bigger interdigitated MSM area. To better understand the saturation behavior of the LAE, we convert pump power to the pump fluence (Fig. 2c). The inset in Fig. 2c highlights that at larger spot sizes, THz power scales up, even at lower fluences. In Fig. 2d we show THz power versus pump area on the LAE at different fluences corresponding to vertical lines in the inset of Fig. 2c. For all fluence values, the THz power scales linearly with the active area. This confirms efficient THz power scaling by simultaneously increasing excitation power and spot size at sufficiently high pulse energy.

We also investigate the effect of bias voltage on the THz emission as a function of pump power. The pump diameter is set to 7 mm to moderate the optical thermal load and ensure 99% of the total optical power is incident on the LAE without clipping. A fully symmetric bipolar rectangular waveform with a 150 kHz modulation frequency, similar to the previous case, is applied as bias voltage. As shown in Fig. 3a, increasing the bias voltage up to 45 $V_{pp}$ leads to higher THz power for all optical powers. Slight saturation at 45 $V_{pp}$ is attributed to thermal load from both optical and electrical sources. For higher voltages up to 49 $V_{pp}$ additional saturation mechanisms reduce the THz emission.

at slightly lower electrical bias fields. Overall, the observed saturation behavior highlights the impact of thermal load from both optical and electrical sources on emitter performance. Designing LAEs with higher resistance could reduce the average current density and thermal load from electrical power [12]. We note however thanks to our higher energy single pulses compared to [15], we still achieve high conversion efficiency and power.

It should be noted that less than 20% of the incident power is absorbed by the LAE, due to the metallization structure covering 75% of the emitter surface and Fresnel loss at the air-GaAs interface [19]. In principle, lens arrays focusing the optical pulses only on the PC areas could address this issue [23]. Another limiting factor for this experiment is the LAE excitation at 515 nm (2.4 eV) largely exceeds the bandgap energy of GaAs (1.42 eV), causing excess photon energy contributes to the heating and also photogenerated carriers are transferred with higher probability to the L-valley [18]. This issue can be addressed by using PC materials optimized for direct excitation by 1 μm lasers, such as ErAs:InAlGaAs, which has been successfully implemented in PC receivers [24], could be a material of choice for future developments.

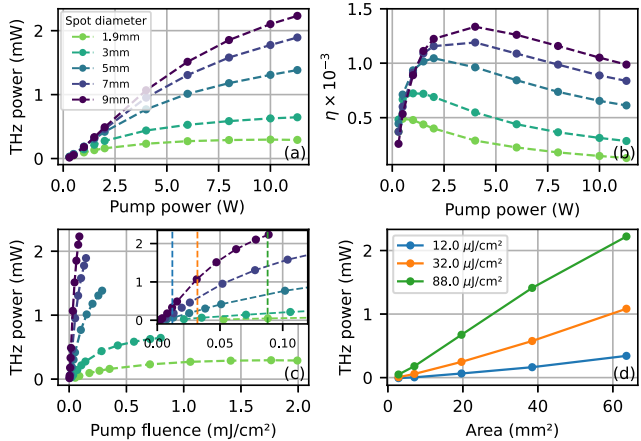

Fig. 2. (a) THz power versus pump power for varying spot sizes. (b) NIR-to-THz conversion efficiency for varying pump diameters. (c) THz power versus pump fluences as a function of pump diameter. (d) THz power versus pump area on the LAE at various fluences. Inset: Zoom into the data.

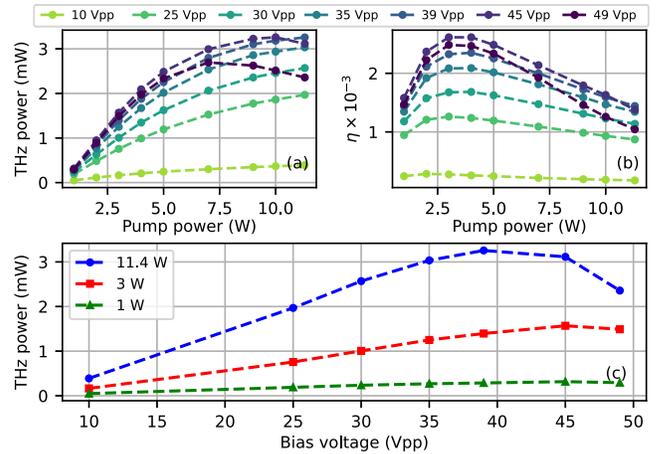

Fig. 3. (a) THz power versus pump power as a function of bias voltage. (b) THz conversion efficiency versus pump power as a function of bias voltage. (c) THz power at fixed optical powers of 11.4 W, 3 W and 1 W versus bias voltage, extracted from (a).

The conversion efficiency peaks at optical powers of ~ 3 W for most of bias voltages, indicating the optimal optical power for LAE excitation. Fig. 3c shows THz power versus bias voltage for three optical pump powers as extracted from Fig. 3b. For the optical power levels of 1 W and 3 W the behavior is very similar. Here an additional saturation mechanism is visible for voltages above 45 $V_{pp}$. In agreement with earlier studies [15], this points to an increase of the average electron effective mass due scattering of carriers into the L-valley. At 11.4 W, the onset for this mechanism is lowered to 40 $V_{pp}$, indicating that under strong heating conditions carriers are transferred to the side valleys already

Fig. 4a presents the averaged THz time-domain trace measured by EOS and its power spectrum. For this measurement the bias voltage is set to 40 $V_{pp}$ at 150 kHz of modulation frequency and excitation power of 11.4 W with 9 mm pump diameter on the LAE. A dark trace is recorded by switching-off the bias voltage but maintaining all other settings, providing insight into the frequency-dependent noise profile of the THz-TDS system.

The dark trace reveals a bandwidth of 3 THz and peak dynamic range of 107 dB in a measurement time of 70 s (1565 traces), as shown in Fig. 4b. These exceptionally high values are rarely reported with EOS, showing the potential of our source for high DR in the future using PC receivers adapted for 1 μm wavelength, which were not available for this experiment. The moderate bandwidth of single cycle THz trace is primarily due to laser's long pulse duration and the larger electrode separation compared to the LAE in [15,16,25]. However, the high DR allows still to fully benefit from the available bandwidth in applications, up to >2 THz.

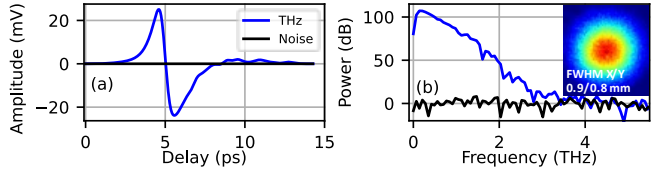

Fig. 4. (a) THz electric field in the time domain. (b) Corresponding THz spectrum in logarithmic scale. Inset, THz beam profile.

The maximum THz power of 6.7 mW is archived by 14.5 mm of pump diameter, 11.4 W optical power, and 55 $V_{pp}$ bipolar square bias voltage. With this spot size, less than 69% of the pump power covers the 1 $cm^2$ square structure, yielding a calculated conversion efficiency of $4.2 \times 10^{-3}$. The improved efficiency with larger spot sizes can be attributed to more homogeneous illumination and better cooling compared to similar excitation levels at higher repetition rates [18]. Therefore, illumination of the LAE with a flat top instead of a Gaussian beam can improve THz conversion efficiency. It is fair to mention that even higher THz power reached ~8 mW with a 14.5 mm spot size on the emitter. However, repeating the measurement at high-voltage caused slight degradation of the emitter due to significant thermal load from high currents and voltages, eventually stabilizing at a reproducible THz power of 6.7 mW, which we report here as record value. This issue could be addressed by incorporating reflection geometry for efficient back-side cooling and improving LAE manufacturing process.

The inset in Fig. 4b shows the THz beam profile recorded by a THz camera (Rigi S2, Swiss Terahertz) with a focused beam FWHM diameter of 0.8 μm × 0.9 μm from a 2D Gaussian fitting, exhibiting a close to ideal Gaussian beam profile. This makes this source a good candidate for application like beam shaping or microscopy which needs a good beam profile and tight focusing. The THz electric field is estimated via similar approach in [26] via integrating the square of electric field over the whole EOS trace in time domain. An electric field of 31 kV/cm at the focus for the maximum THz power of 6.7 mW. With a 1:4 imaging system, instead of current 1:2, a focus spot of 0.57 μm × 0.72 μm at FWHM is also achieved.

In summary, we demonstrate average power scaling of a microstructured large-area emitter excited by a frequency-doubled, widely available Yb-laser amplifier. At 11.4 W of excitation power, we achieve high THz average power of 6.7 mW at 400 kHz repetition rate, reaching the highest average power so far demonstrated with PC emitters. EOS traces exhibit a high dynamic range of 107 dB in 70 s integration time and a spectral bandwidth extending up to 3 THz. Furthermore, we present what are the critical challenges identified for future scaling. Further thermal management challenges arising from both optical and electrical dissipation are limiting, however the high individual driving pulse energy allows to make efficient use of the maximum area of the emitter, still allowing for high conversion efficiency. In future realizations, LAEs based on other materials with higher resistivity, suitable for direct excitation at 1030 nm, in reflection geometry to improve cooling efficiency and using microlens arrays to focused optical pulse only to the photoconductive area could significantly improve the current results towards the 100-mW range. The demonstrated source represents an attractive, widely accessible option for the generation of high average power, broadband THz radiation.

**Funding.** Ruhr-Universität Bochum (Open Access Publication Funds); Deutsche Forschungsgemeinschaft (287022738 TRR 196, SFB/TRR 196 , for Project M01.b).

**Acknowledgments**. The research was carried out in the Research Center Chemical Sciences and Sustainability of the University Alliance Ruhr at Ruhr-University Bochum.

**Disclosures.** The authors declare no conflicts of interest.

**Data availability.** Data underlying the results presented in this paper are available in Zenodo.

**References**
1. M. Tonouchi, Nature Photon **1**, 97 (2007).
2. B. Ferguson and X.-C. Zhang, Nature Mater **1**, 26 (2002).
3. T. L. Cocker, D. Peller, P. Yu, et al., Nature **539**, 263 (2016).
4. H. Lin, P. Braeuninger-Weimer, V. S. Kamboj, et al., Sci. Rep. **7**, 10625 (2017).
5. T. Vogel and C. J. Saraceno, J Infrared Milli Terahz Waves (2024).
6. J. Buldt, H. Stark, M. Müller, et al., Opt. Lett. **46**, 5256 (2021).
7. T. Vogel, S. Mansourzadeh, and C. J. Saraceno, Opt. Lett. **49**, 4517 (2024).
8. D. R. Bacon, J. Madéo, and K. M. Dani, J. Opt. **23**, 064001 (2021).
9. P. Chen, M. Hosseini, and A. Babakhani, Micromachines **10**, 367 (2019).
10. R. B. Kohlhaas, S. Breuer, L. Liebermeister, et al., Appl. Phys. Lett. **117**, 131105 (2020).
11. R. B. Kohlhaas, L. Gingras, E. Dardanis, et al., in *IRMMW-THz* (IEEE, 2022), pp. 1–2.
12. A. Dohms, N. Vieweg, S. Breuer, et al., IEEE Trans. THz Sci. Technol. **14**, 857 (2024).
13. E. Budiarto, J. Margolies, S. Jeong, et al., IEEE J. Quantum Electron. **32**, 1839 (1996).
14. S. Winnerl, J. Infrared Millim. Terahertz Waves **33**, 431 (2012).
15. M. Beck, H. Schäfer, G. Klatt, et al., Opt. Express **18**, 9251 (2010).
16. N. Nilforoushan, C. Kidd, A. Fournier, et al., Appl. Phys. Lett. **123**, 241107 (2023).
17. A. Dreyhaupt, S. Winnerl, M. Helm, et al., Opt. Lett. **31**, 1546 (2006).
18. M. Khalili, T. Vogel, Y. Wang, et al., Opt. Express (2024).
19. A. Dreyhaupt, S. Winnerl, T. Dekorsy, et al., Appl. Phys. Lett. **86**, 121114 (2005).
20. B. B. Hu, J. T. Darrow, X.-C. Zhang, et al., Appl. Phys. Lett. **56**, 886 (1990).
21. Q. Wu and X. -C. Zhang, Appl. Phys. Lett. **67**, 3523 (1995).
22. A. Leitenstorfer, Physica B: Condensed Matter **272**, 348 (1999).
23. A. Garufo, P. M. Sberna, G. Carluccio, et al., IEEE Trans. THz Sci. Technol. **9**, 221 (2019).
24. T. Vogel, S. Mansourzadeh, U. Nandi, et al., IEEE Trans. THz Sci. Technol. **14**, 139 (2024).
25. J. Madéo, N. Jukam, D. Oustinov, et al., Electron. Lett. **46**, 611 (2010).
26. M. Clerici, M. Peccianti, B. E. Schmidt, et al., Phys. Rev. Lett. **110**, 253901 (2013).

## References (Long)

1. M. Tonouchi, "Cutting-edge terahertz technology," Nature Photon **1**, 97–105 (2007).
2. B. Ferguson and X.-C. Zhang, "Materials for terahertz science and technology," Nature Mater **1**, 26–33 (2002).
3. T. L. Cocker, D. Peller, P. Yu, J. Repp, and R. Huber, "Tracking the ultrafast motion of a single molecule by femtosecond orbital imaging," Nature **539**, 263–267 (2016).
4. H. Lin, P. Braeuninger-Weimer, V. S. Kamboj, D. S. Jessop, R. Degl'Innocenti, H. E. Beere, D. A. Ritchie, J. A. Zeitler, and S. Hofmann, "Contactless graphene conductivity mapping on a wide range of substrates with terahertz time-domain reflection spectroscopy," Sci. Rep. **7**, 10625 (2017).
5. T. Vogel and C. J. Saraceno, "Advanced Data Processing of THz-Time Domain Spectroscopy Data with Sinusoidally Moving Delay Lines," J Infrared Milli Terahz Waves (2024).
6. J. Buldt, H. Stark, M. Müller, C. Grebing, C. Jauregui, and J. Limpert, "Gas-plasma-based generation of broadband terahertz radiation with 640 mW average power," Opt. Lett. **46**, 5256 (2021).
7. T. Vogel, S. Mansourzadeh, and C. J. Saraceno, "Single-cycle, 643 mW average power terahertz source based on tilted pulse front in lithium niobate," Opt. Lett. **49**, 4517 (2024).
8. D. R. Bacon, J. Madéo, and K. M. Dani, "Photoconductive emitters for pulsed terahertz generation," J. Opt. **23**, 064001 (2021).
9. P. Chen, M. Hosseini, and A. Babakhani, "An Integrated Germanium-Based THz Impulse Radiator with an Optical Waveguide Coupled Photoconductive Switch in Silicon," Micromachines **10**, 367 (2019).
10. R. B. Kohlhaas, S. Breuer, L. Liebermeister, S. Nellen, M. Deumer, M. Schell, M. P. Semtsiv, W. T. Masselink, and B. Globisch, "637 µW emitted terahertz power from photoconductive antennas based on rhodium doped InGaAs," Appl. Phys. Lett. **117**, 131105 (2020).
11. R. B. Kohlhaas, L. Gingras, E. Dardanis, R. Holzwarth, S. Breuer, M. Schell, and B. Globisch, "Fiber-coupled THz TDS system with mW-level THz power," in *IRMMW-THz* (IEEE, 2022), pp. 1–2.
12. A. Dohms, N. Vieweg, S. Breuer, T. Heßelmann, R. Herda, N. Regner, S. Keyvaninia, M. Gruner, L. Liebermeister, M. Schell, and R. B. Kohlhaas, "Fiber-Coupled THz TDS System With mW-Level THz Power and up to 137-dB Dynamic Range," IEEE Trans. THz Sci. Technol. **14**, 857–864 (2024).
13. E. Budiarto, J. Margolies, S. Jeong, J. Son, and J. Bokor, "High-intensity terahertz pulses at 1-kHz repetition rate," IEEE J. Quantum Electron. **32**, 1839–1846 (1996).
14. S. Winnerl, "Scalable Microstructured Photoconductive Terahertz Emitters," J. Infrared Millim. Terahertz Waves **33**, 431–454 (2012).
15. M. Beck, H. Schäfer, G. Klatt, J. Demsar, S. Winnerl, M. Helm, and T. Dekorsy, "Impulsive terahertz radiation with high electric fields from an amplifier-driven large-area photoconductive antenna," Opt. Express **18**, 9251 (2010).
16. N. Nilforoushan, C. Kidd, A. Fournier, J. Palomo, J. Tignon, S. Dhillon, E. Lhuillier, L. Li, A. G. Davies, E. H. Linfield, J. R. Freeman, and J. Mangeney, "Efficient THz generation from low-temperature-grown GaAs photoconductive antennas driven by Yb-doped fiber amplifier at 200 kHz repetition rate," Appl. Phys. Lett. **123**, 241107 (2023).
17. A. Dreyhaupt, S. Winnerl, M. Helm, and T. Dekorsy, "Optimum excitation conditions for the generation of high-electric-field terahertz radiation from an oscillator-driven photoconductive device," Opt. Lett. **31**, 1546 (2006).
18. M. Khalili, T. Vogel, Y. Wang, S. Mansourzadeh Ashkani, A. Singh, S. Winnerl, and C. Saraceno, "Microstructured large-area photoconductive terahertz emitters driven at high average power," Opt. Express (2024).
19. A. Dreyhaupt, S. Winnerl, T. Dekorsy, and M. Helm, "High-intensity terahertz radiation from a microstructured large-area photoconductor," Appl. Phys. Lett. **86**, 121114 (2005).
20. B. B. Hu, J. T. Darrow, X.-C. Zhang, D. H. Auston, and P. R. Smith, "Optically steerable photoconducting antennas," Appl. Phys. Lett. **56**, 886–888 (1990).
21. Q. Wu and X.-C. Zhang, "Free-space electro-optic sampling of terahertz beams," Appl. Phys. Lett. **67**, 3523–3525 (1995).
22. A. Leitenstorfer, "Ultrafast high-field transport in semiconductors," Physica B: Condensed Matter **272**, 348–352 (1999).
23. A. Garufo, P. M. Sberna, G. Carluccio, J. R. Freeman, D. R. Bacon, L. Li, J. Bueno, J. J. A. Baselmans, E. H. Linfield, A. G. Davies, N. Llombart, and A. Neto, "A Connected Array of Coherent Photoconductive Pulsed Sources to Generate mW Average Power in the Submillimeter Wavelength Band," IEEE Trans. THz Sci. Technol. **9**, 221–236 (2019).
24. T. Vogel, S. Mansourzadeh, U. Nandi, J. Norman, S. Preu, and C. J. Saraceno, "Performance of Photoconductive Receivers at 1030 nm Excited by High Average Power THz Pulses," IEEE Trans. THz Sci. Technol. **14**, 139–151 (2024).
25. J. Madéo, N. Jukam, D. Oustinov, M. Rosticher, R. Rungsawang, J. Tignon, and S. S. Dhillon, "Frequency tunable terahertz interdigitated photoconductive antennas," Electron. Lett. **46**, 611–613 (2010).
26. M. Clerici, M. Peccianti, B. E. Schmidt, L. Caspani, M. Shalaby, M. Giguère, A. Lotti, A. Couairon, F. Légaré, T. Ozaki, D. Faccio, and R. Morandotti, "Wavelength Scaling of Terahertz Generation by Gas Ionization," Phys. Rev. Lett. **110**, 253901 (2013).